\documentclass[twocolumn,showpacs,amsmath,amssymb]{revtex4}
\usepackage{dcolumn}
\usepackage{bm}

\usepackage[pdftex]{graphicx,color}

\newcommand\Dfrtl[1]{\ensuremath{\,\mathrm{d}#1\,}}

\begin{document}


\title{
Feshbach modulation spectroscopy
}

\date{\today}
\author{Andreas Dirks}
\email{andreas@physics.georgetown.edu}
\affiliation{Department of Physics, Georgetown University, Washington, DC 20057, USA}
\author{Karlis Mikelsons}
\affiliation{Department of Physics, Georgetown University, Washington, DC 20057, USA}
\author{H. R. Krishnamurthy}
\affiliation{Centre for Condensed Matter Theory, Department of Physics,\\
Indian Institute of Science, Bangalore 560012, India, and \\
Jawaharlal Nehru Centre for Advanced Scientific Research, Bangalore 560064, India
}
\author{J. K. Freericks}
\affiliation{Department of Physics, Georgetown University, Washington, DC 20057, USA}

\begin{abstract}
In the vicinity of a Feshbach resonance, a system of ultracold atoms on an optical lattice undergoes
rich physical transformations which involve molecule formation and hopping of molecules on the lattice and 
thus goes beyond a single-band Hubbard model description. We propose to probe the behavior 
of this system with a harmonic modulation of the magnetic field, and thus of
the scattering length,
\emph{across} the Feshbach resonance, as an alternative to lattice-depth modulation spectroscopy.
In the regime in which the single-band Hubbard model is still valid, we provide simulation data for this 
type of spectroscopy. The method may uncover a route towards the efficient creation of ultracold molecules
and provides an alternate means for lattice modulation spectroscopy.
\end{abstract}

\maketitle

The field of ultracold atoms in optical lattices has been opening up new possibilities which include a
controlled experimental realization of the fermionic Hubbard model \cite{blochreview, Esslinger2010}. Further challenges and opportunities arise with the 
idea of manipulating and controlling molecules on an optical lattice.
Molecules allow a much wider range of physical phenomena to be modelled and
studied than is possible with atoms.
However, it is more difficult to cool molecules down to the ground state via laser cooling, due to their more complex level structure which includes rotational
and vibrational degrees of freedom.
The cooling of individual atoms to a very low temperature followed by the formation of so-called preformed molecules on 
the optical lattice is thus a promising alternative \cite{Freericks2010}.
Near a Feshbach resonance, bound states of these preformed molecules occur. Depending on the value of the magnetic field, molecules 
form and hop from one lattice site to the other; these processes are governed by the complex Fermi Resonance Hamiltonian (FRH) \cite{Wall2012}.

It is crucial to understand the FRH physics in order to control and optimize the formation process. 
Experimentally, the understanding may be facilitated by a spectroscopic method which we propose in the following. 
In case of the single-band Hubbard model, the so-called lattice modulation spectroscopy has been proven useful to study non-equilibrium dynamics.
In lattice modulation spectroscopy, the intensity of the laser defining the optical lattice is varied harmonically. As a result, the hopping amplitude 
and the interaction strength \emph{both} change as a function of time and the Mott gap can be measured directly in the experiment.

Working with a modulated magnetic field near a Feshbach resonance in order to
examine a modulation of the scattering length has been
investigated in a number of different contexts. It originally was used to
describe Feshbach resonance management~\cite{kevrikidis_2003,abdullaev_2003}
which controlled ``breathers'' and solitons in trapped bosonic systems. Next,
it was used to show how many-body effects and the periodic driving could push
the tunneling to vanish~\cite{hanggi_2009} also in bosonic systems. More
recently, it has been used to illustrate how one can obtain correlated
hopping in bosonic systems when the amplitude of the magnetic field
oscillation is small~\cite{rapp_2012,liberto_2014,greshner_2014}. Experiments
have also been carried out for bosonic
systems~\cite{pollack_2010} to examine driven collective excitations. Here,
we focus on the Fermi version of the Hubbard model,
and examine situations where the driving is pushed much closer to the
Feshbach resonance, and even passing through it, where nonlinear effects
become important.

While it is a powerful method to probe the atomic Hubbard physics, lattice modulation spectroscopy  does not modify the sign of the interaction 
strength and is thus fundamentally limited when more general physics issues
such as the molecule formation are to be studied.

We thus propose to probe the system with a harmonic modulation of the magnetic field
\begin{equation}
B(t) = \bar B + \chi_{[0,t_\text{mod}]}(t) \cdot \Delta B \sin \omega t,
\label{eq:fieldmod}
\end{equation}
near the Feshbach resonance, where
\begin{equation}
\chi_I(t) =
\begin{cases}
1 & \text{if } t\in I \\
0 & \text{otherwise}
\end{cases}
\end{equation}
is the characteristic function of the modulation interval.
In order to provide some numerical data for this spectroscopy method, we consider a system of
fermionic $^{40}$K atoms subject to the $ab$-Feshbach resonance \cite{Chin2010} 
in an optical lattice with a laser wavelength of $1064\,$nm.

While at present, providing numerical results for the full FRH is beyond
reach, we present numerical data for the atomic Hubbard limit which are valid in
the early stages of the preformed molecule formation process: \cite{Hubbard1963}
\begin{equation}
\begin{split}
H(t) =&\, -J(t) \sum_{\langle
i,j\rangle,\sigma}\left(c^\dagger_{i\sigma}c_{j\sigma} +\text{h.c.} \right) 
\\&+ U(t) \sum_i n_{i\uparrow} n_{i\downarrow}+\sum_{i\sigma} \epsilon_i n_{i\sigma} .
\end{split}
\label{eq:Hubbard}
\end{equation}

The time dependence of the lattice parameters reads $J(t)=J_0 = \text{const}$ and $U(t) = g(t) \int |w(\vec r)|^4 d^3r$, where $w(\vec r)$ is the maximally localized Wannier function \cite{Kohn1959}. 
The time-dependent coupling constant $g(t)=4\pi\hbar^2a(t)/m$ is determined by the mass $m$ 
of the $^{40}$K atoms and the $s$-wave scattering length 
\begin{equation}
a(t) = a_{bg} \left( 1 - \frac{\Delta}{B(t)-B_0} \right),
\end{equation}
where $a_{bg}=174\,a_0$ is the background scattering, $B_0=202.1\,G$ is the position of the Feshbach resonance and $\Delta = 8.0\,G$ is its width.

For simplicity, we consider a translationally invariant lattice in three dimensions at half filling in the Mott-insulating phase and
study the behavior of the double occupancy.
With a higher double occupancy, molecule formation is more likely to occur in the later stages of the driving of the full FRH system.
Computationally, we employ a strong-coupling
approach which works well at finite temperatures larger than the hopping and has already successfully modelled the conventional modulation spectroscopy \cite{Mikelsons2012, dirks2013, dirks2014}.
In order to ensure the accuracy of the approach, we constrain the studied
parameter range to a maximum value $j_\text{max} := \text{max}\,\{J_0/U(t)\}_{t\in\mathbb{R}}\approx 1/24$.

\begin{figure}
\includegraphics[width=\linewidth]{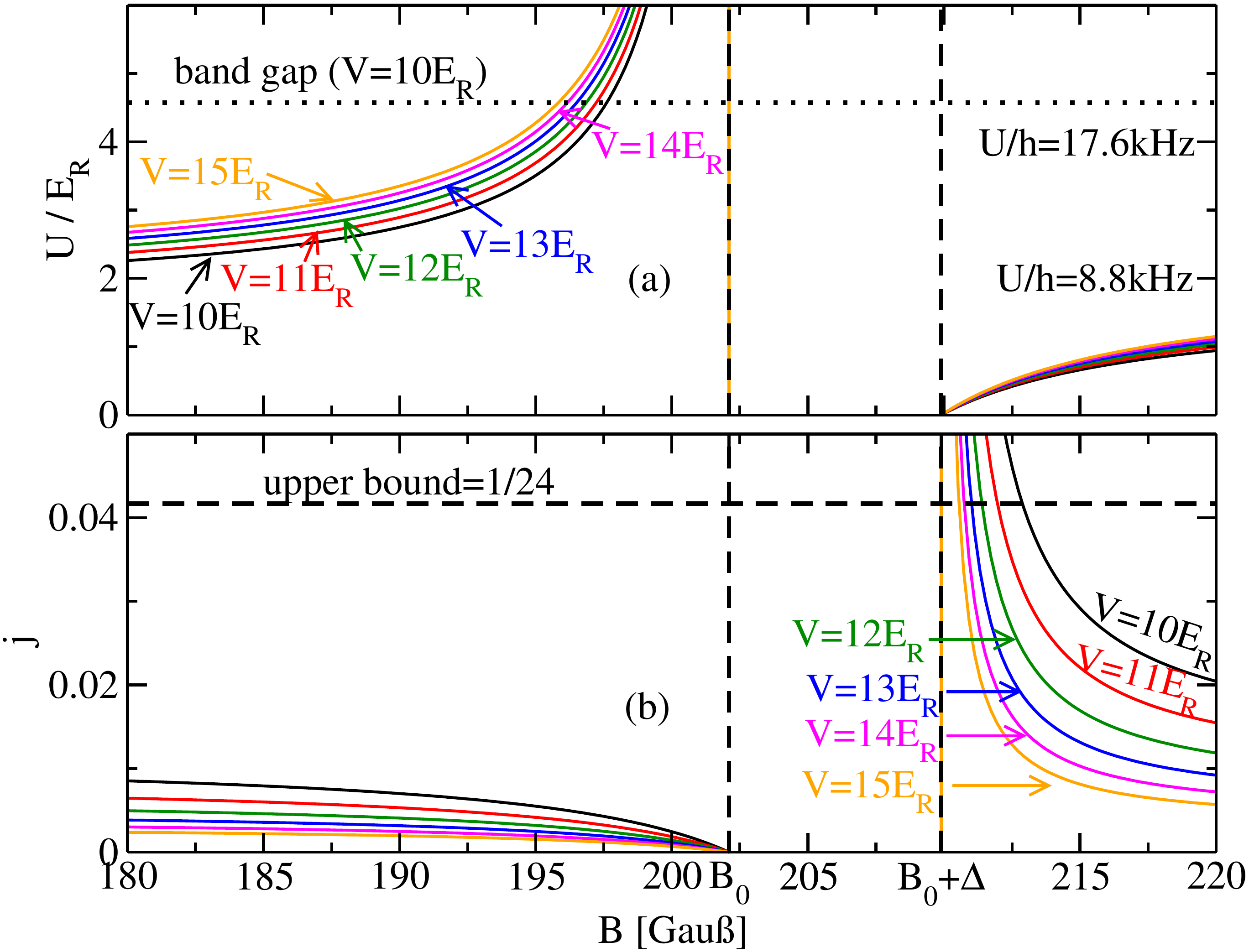}
\caption{(Color online) $ab$-Feshbach resonances of the system $^{40}$K for different lattice depths. The upper panel (a) shows the dependence of the interaction on the magnetic field and
the lower panel (b) shows the resulting renormalized hopping $j=J_0/U(B)$.
}
\label{fig:mapsFB}
\end{figure}

For each lattice depth, the Feshbach resonance has a different effect on the hopping relative to the interaction, i.e.~on $j(t):=J_0/U(t)$.
Also, the magnetic field dependence of the hopping strength in units of the interaction $j(B):=J_0/U(B)$ plays a key role in the 
Feshbach spectroscopy of the Hubbard model. In Fig.~\ref{fig:mapsFB}, panel (b) shows this map for several lattice depths. 
Panel (a) shows the corresponding interaction strength.
We limit our consideration to the interval $[0,j_\text{max}]$ indicated by the dashed line in panel (b). In addition, we assume that the 
amplitude of the magnetic field is realistically smaller than 5~G for the necessary modulation
frequency of a couple of kHz. 
We also require the interaction to be significantly lower than the non-interacting bandgap which is also displayed in panel (a) at lattice depth $V=10E_R$.
This, together with the requirement that the renormalized hopping $j$, while
small, should be large enough for the effects due to changes in it arising
from changes in $B$ to be measurable, 
constrains the considered parameter range to the right branches displayed in panel (b).
Thus we consider magnetic field values within the interval $(B_0 + \Delta,
220G]$
and lattice depths equal to or larger than $10E_R$ (for
smaller $V$, the bandgap to the second band would be too small).

In experiments, the upper bound for the hopping does not apply. However, in the vicinity of the resonance, the strong dependence of the 
effective hopping on the field also results in a stronger dependence on the inhomgeneities of the magnetic field. It is thus also reasonable 
to keep the value of $j$ below a certain threshold in experiments to reduce the effects of inhomogeneity.

In addition to the mean value of the magnetic field, other important parameters to be considered are the amplitude
and the frequency of the field modulation. If the physical response of the system is sensitive to these values, this may help to determine 
unknown model properties (such as the lattice depth in the experiments) more precisely than possible in lattice depth modulation spectroscopy. 
In order to study such effects, we investigate the frequency dependence of the doublon production rate for fixed windows of magnetic field modulation.

\begin{figure*}
\includegraphics[width=\linewidth]{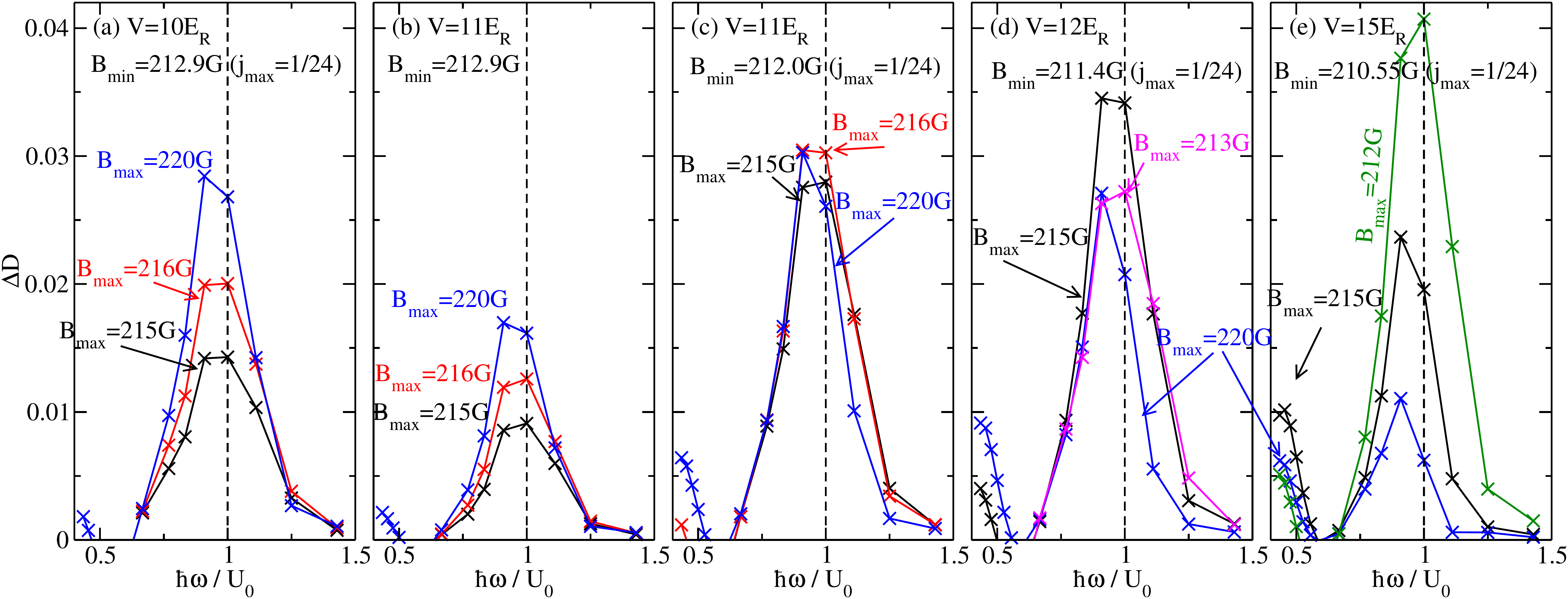}
\caption{(Color online) Doublon production at different values of
$B_\text{max}$. Each panel represents different values of $V$ and $B_\text{min}$.
In panels (a), (c),(d), and (e), $B_\text{min}$ is chosen such that
$j_\text{max}=1/24$. The initial temperature is $k_BT=0.1U_0$.}
\label{fig:widefig}
\end{figure*}

The field modulation in Eq.~\eqref{eq:fieldmod} is parametrized by the magnetic field amplitude $\Delta B$, the average field value $\bar B$, the length of the 
modulation time interval $t_\text{mod}$, and the modulation frequency
$\omega$. $\Delta B$ and $\bar B$ can alteratively be expressed in terms of
the minimum 
and maximum values of the field strength, $B_\text{min}=\bar B-\Delta B$ and $B_\text{max} = \bar B + \Delta B$. These values also determine the minimum and 
maximum values of the renormalized hopping $j(B)=J_0/U(B)$. In order to translate $B_\text{min/max}$ into $j_\text{min/max}$ one uses Fig.~\ref{fig:mapsFB}(b).

We consider three field modulation intervals $[B_\text{min}, B_\text{max}]$ first and compare the behavior for two
lattice depths. Depending on the frequency, the field is modulated over a time interval $[0,t_\text{max}]$, with
\begin{equation}
t_\text{max}(\omega) = \left\lfloor \tilde t_\text{max} \cdot \left(\frac{2\pi}{\omega}\right)^{-1} \right\rfloor \times \frac{2\pi}{\omega},
\end{equation}
and $\tilde t_\text{max}\times U_0/\hbar = 29$, resulting in 2 to 6 field
modulation cycles for $\hbar \omega / U_0 = 0.5 \dots 1.5$, where
$U_0:=U(\bar B)$.
As a physical observable, we study the excitation from the lower to the upper
Hubbard band which is measured by the double occupancy per site
\begin{equation}
D(t) = \langle n_{\uparrow} n_{\downarrow} \rangle (t)
\end{equation}
and study the increase in this quantity, which we measure as
\begin{equation}
\Delta D := \frac{U_0}{h}\int_{\tilde t_\text{max}+2-h/U_0}^{\tilde t_\text{max}+2}\Dfrtl t D(t) - D(t_0).
\end{equation}
That is, the end value has been averaged over one oscillation period of a
resonantly excited Hubbard system and compared to the initial value.

Figure \ref{fig:widefig} shows the resulting frequency dependence of $\Delta D$ for three different values of $B_\text{max}$,
while we keep the minimum field value constant at $B_\text{min}= 212.9G$. 
Panels (a) and (b) show the results for the lattice depths $V=10\,E_R$ and $V=11\,E_R$, respectively.
Since the hopping $j_\text{max}$ is smaller for a deeper lattice, less doublons are produced for $V=11E_R$ than for $V=10E_R$. However, the behavior of the curves as a
function of $B_\text{max}$ is qualitatively the same for the two lattice depths.

Hence we discuss the dependence of the resonance curves on $B_\text{min}$ in more detail.
Figure \ref{fig:widefig} shows several resonance curves for two slightly
different values of $B_\text{min}$ in panels (b) and (c), respectively. It shows that 
even the qualitative behavior of the Feshbach modulation can be quite sensitive to the details of the model. 
In panel (c), the shape and the strength of the resonances
are approximately the same. For the slightly larger value of $B_\text{min}$ shown in panel (b), the resonance curves change drastically as a function of 
$B_\text{max}$. The reason for this qualitatively different behavior is that in case (c) a larger fraction of the steep portion of the renormalized hopping $j$ as a function 
of $B$ (see Fig.~\ref{fig:mapsFB}) is sampled in the modulation procedure than in case (b). 
An effect which both the cases (b) and (c) have in common is that the maximum in doublon production is shifted towards smaller frequencies for larger values of $B_\text{max}$.
The reason for this may be the lower time-averaged value of the interaction strength for larger values of $B_\text{max}$
in units of the respective values for $U_0=U(\bar B)$. For example, in the
simplified case $B_\text{min}=B_0+\Delta$, the time-averaged
value of the interaction $U_\text{tavg}$ can be approximately written as 
$U_\text{tavg}/U_0=1-(U_\text{bg}/2U_0)\times b^2$, where $U_\text{bg}$ is the interaction associated with
the background scattering $a_\text{bg}$ and $b = (B_\text{max} -
B_\text{min})/2\Delta$.
A similar relation can be derived for the more realistic $B_\text{min}>B_0+\Delta$.
However, since the width of the resonance is almost constant in both panels (b) and (c), this reasoning cannot be the whole
story.

Furthermore, we can also compare the resonance curves for several lattice
depths at a fixed maximum value $j_\text{max}$ of the renormalized hopping. This corresponds to identifying the value of 
$B_\text{min}$ for which the value $j_\text{max}$ is obtained, for each lattice depth. In this case, we choose $j_\text{max} = 1/24$, which is also the upper theoretical bound we introduced 
previously in order to assure the convergence of the strong-coupling method.
Panels (a), (c),(d), and (e) of Fig.~\ref{fig:widefig} show data for
different lattice depths at a constant maximum value of $j$.
We again find that the dependence on $B_\text{max}$ may depend very much on the 
lattice depth. While for the shallow lattice, $V=10E_R$, increasing the modulation amplitude yields a stronger signal, we observe the \emph{opposite} effect in a deeper lattice, $V=15E_R$.
This striking difference is due to the increasing nonlinearity of $j(B)$ as
$V$ increases. For a shallow lattice, $j(B)$ still exhibits a nearly linear
behaviour, so the peak strength is proportional to the amplitude.
In a deep lattice, $j(B)$ is strongly nonlinear and the system is rather
kicked than driven. An increased amplitude decreases the kick strength in a
deep lattice, because $j$ is close to $j_\text{max}$ for shorter time spans
during the modulation. 
As the lattice gets deeper, a second order peak appears at $\hbar\omega = U_0/2$, which is approximately as strong as the first-order peak for the strong modulation amplitude.
The lattice depths between $V=10E_R$ and $V=15E_R$ interpolate between these
two behaviors. In the very deep lattice, for $V=15E_R$, the strongest doublon production can be achieved with a rather small amplitude
corresponding to $B_\text{max}=212G$, or $\Delta B \approx 0.73G$.

\begin{figure}
\includegraphics[width=\linewidth]{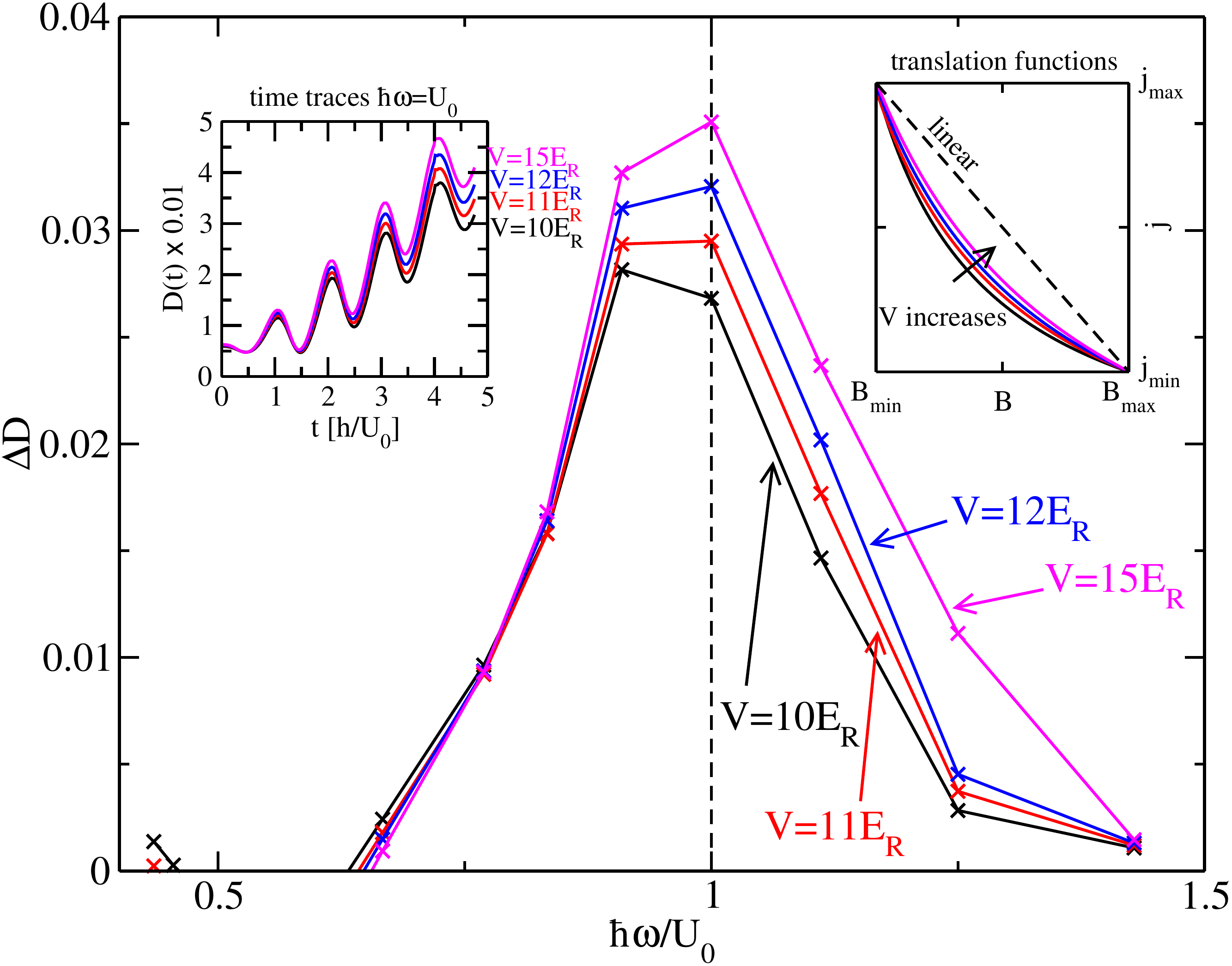}
\caption{(Color online) Magnetic modulation with the renormalized hopping  $j$ oscillating within the interval $[j_\text{min}, j_\text{max}]$, with $j_\text{max}=1/24$ and $j_\text{min}=j_\text{max}/2$ for different lattice depths. The corresponding magnetic field intervals $I_B=[B_\text{min}, B_\text{max}]$ are
$I_B(V=10E_R)=[212.9G, 219.56G]$, $I_B(V=11E_R)=[212.0G,215.52G]$, $I_B(V=12E_R)=[211.4G, 213.59G]$, $I_B(V=15E_R)=[210.55G, 211.32G]$. }
\label{fig:jminjmax}
\end{figure}

Finally, in order to compare different lattice depths, we fix the values of $B_\text{min}$ and $B_\text{max}$ in such a way that the renormalized hopping oscillates between the values $j_\text{min} = 1/48$ and 
$j_\text{max} = 1/24$. The resulting resonance curves at different lattice depths are shown in Fig.~\ref{fig:jminjmax}.
In contrast to the scenarios discussed in Fig.~\ref{fig:widefig}, the curves are now essentially identical. This underlines the central role of the renormalized hopping in interpreting both Feshbach and 
lattice depth modulation spectroscopy. However, we also observe a tendency towards a strong doublon production for deeper lattices. As can be seen in the left inset of Fig.~\ref{fig:jminjmax}, this is not related to the initial number of doubly occupied sites 
which is essentially identical for each lattice depth. Rather, the tendency
is due to the shape of the translation function between magnetic field and effective hopping, as shown in the right inset of Fig.~\ref{fig:jminjmax}.
As the lattice depth is increased, the convexity of the translation function is decreased and approaches a linear behavior. This gives rise to an increase in the doublon production.

\emph{Conclusion.}
In this work we have proposed an experimental technique that is an alternative to conventional lattice modulation spectroscopy,
where tuning and modulating a magnetic field near a Feshbach resonance allows for the system to have a time dependent 
interaction, with a constant hopping (the renormalized hopping, of course is time dependent). This changes the behavior of the 
driving of the system from a more kicked drive in the conventional approach to a smoother evolution in this case. We find that
in some cases, the signal can have strong resonant effects that require fine tuning of the magnetic field, and hence have the potential to produce
higher precision measurements. In addition, we find that the ``two-photon'' peak at a frequency equal to half the average interaction
strength, is often enhanced in these systems making it easier to study nonlinear excitation effects. Finally, we conjecture that even more
interesting behavior will occur when the Feshbach modulation spectroscopy is pushed through the Feshbach resonance itself and allows for molecule
formation. 
The many mutually coupled degrees of freedom in the FRH \cite{Wall2012} promise a rich variety of
physical effects which will be interesting to investigate both experimentally
and theoretically. In particular, it will be interesting to explore the channels
that lead to molecule formation spectroscopically.
We do not yet have the ability to model and calculate the behavior of such spectroscopy, but experiments could potentially investigate
such effects in the near future.

\section {Acknowledgments}
This work was supported by a
MURI grant from the Air Force Office of Scientific Research
numbered FA9559-09-1-0617. Supercomputing resources
came from a challenge grant of the DoD at the Engineering Research
and Development Center and the Air Force Research and
Development Center. The collaboration was supported
by the Indo-US Science and Technology Forum under
the joint center numbered JC-18-2009 (Ultracold atoms).
JKF also acknowledges the McDevitt bequest at Georgetown.
HRK acknowledges support of the Department of 
Science and Technology in India.
AD was in part supported by the Collaborative Research Center 
1073 of the German Research Council.

\end{document}